\documentclass[twocolumn,conference]{IEEEtran}
\usepackage{color}
\usepackage{array}
\usepackage{verbatim}
\usepackage{booktabs}
\usepackage{graphicx}
\usepackage[numbers]{natbib}

\makeatletter

\def\ps@IEEEtitlepagestyle{%
  \def\@oddfoot{\mycopyrightnotice}%
  \def\@evenfoot{}%
}

\newcommand\copyrighttext{%
  \footnotesize \textcopyright~2017 IEEE. Personal use of this material is permitted. Permission from IEEE must be obtained for all other uses, in any current or future media, including reprinting/republishing this material for advertising or promotional purposes, creating new collective works, for resale or redistribution to servers or lists, or reuse of any copyrighted component of this work in other works.}

\def\mycopyrightnotice{%
  {\fbox{\parbox{\dimexpr\textwidth-\fboxsep-\fboxrule\relax}{\copyrighttext}}}
  \gdef\mycopyrightnotice{}
}

\usepackage[font=footnotesize]{subfig}

\usepackage[factor=1100,stretch=10,shrink=20,tracking=true,kerning=true,spacing=true]{microtype}

\usepackage{textcomp}

\usepackage {url}

\def\baselinestretch{0.95}
\usepackage[font=small]{caption}
\usepackage{paralist}

\makeatother

\begin{document}

\title{Automating Ethernet VPN Deployment in SDN-based Data Centers}

\author{\IEEEauthorblockN{Kyoomars Alizadeh Noghani\IEEEauthorrefmark{1}, Cristian Hernandez Benet\IEEEauthorrefmark{1}, Andreas Kassler\IEEEauthorrefmark{1}\\
\vspace{-1ex}
Antonio Marotta\IEEEauthorrefmark{1}, Patrick Jestin\IEEEauthorrefmark{2}, Vivek V. Srivastava\IEEEauthorrefmark{2}}
\\
\IEEEauthorblockA{\IEEEauthorrefmark{1} Karlstad University, \IEEEauthorrefmark{2} Ericsson AB 
\vspace{-2ex}
\\\\
\{kyoomars.noghani-alizadeh, cristian.hernandez-benet, andreas.kassler, antonio.marotta\}@kau.se
\\\vspace{-2ex}
\{patrick.jestin, vivek.v.srivastava\}@ericsson.com}}

\maketitle

\begin{abstract}
Layer 2 Virtual Private Network (L2VPN) is widely deployed in both service provider networks and enterprises. However, legacy L2VPN solutions have scalability limitations in the context of Data Center (DC) interconnection and networking which require new approaches that address the requirements of service providers for virtual private cloud services. Recently, Ethernet VPN (EVPN) has been proposed to address many of those concerns and vendors started to deploy EVPN based solutions in DC edge routers. 
However, manual configuration leads to a time-consuming, error-prone configuration and high operational costs. Automating the EVPN deployment from cloud platforms such as OpenStack enhances both the deployment and flexibility of EVPN Instances (EVIs). This paper proposes a Software Defined Network (SDN) based framework that automates the EVPN deployment and management inside SDN-based DCs using OpenStack and OpenDaylight (ODL). We implemented and extended several modules inside ODL controller to manage and interact with EVIs and an interface to OpenStack that allows the deployment and configuration of EVIs. We conclude with scalability analysis of our solution.
\end{abstract}


\section{Introduction}
\label{sec:intro}
Virtual Private Network (VPN) technology is widely used to interconnect geographically distributed sites. Among VPN solutions, Layer-2 VPN (L2VPN) has been evolved and attracted significant interest over recent years due to its flexibility and transparency requirements. Additionally, several applications that run in Virtual Machines (VMs) inside virtual Data Centers (DCs) require L2 connectivity which cannot be simply replaced with L3 solutions. Traditionally, Virtual Private Lan Service (VPLS)~\cite{rfc4761} has been adopted as the best L2VPN solutions for DC interconnection since its ability to span VLANs between different sites, enabling the extension of customer VLANs towards DCs. However, VPLS has its limitations in terms of redundancy, scalability, flexibility, and limited forwarding policies. Additionally, Internet Service Providers (ISPs) typically use Multiprotocol Label Switching (MPLS) to interconnect DCs given its flexibility and ease of deployment, and it is important that the VPN service is designed to function upon MPLS technology. To address the aforementioned problems, Ethernet VPN (EVPN)~\cite{rfc7432} has been proposed which allows to create flexible L2 interconnect solutions over MPLS.

On-demand cloud services create network and orchestration requirements such as to deploy and destroy VMs and provide them network connectivity across the DCs as quickly as possible. In order to achieve this goal, ISP and DC administrators need to address two main problems. The first focuses on providing a flexible network management automation. Despite the efforts made to create protocols such as Network Configuration Protocol (NETCONF)~\cite{rfc6241} and SNMP~\cite{rfc1157} aiming to offer a faster configuration of network devices, these do not allow the ISP to deal with on-demand services and does not provide enough flexibility due to its vendor-dependency. The introduction of a new customer or service involves a set of configuration procedures which involves ISPs to go through a time-consuming and error-prone configuration process. The second aim is to reduce the control plane complexity of MPLS-based VPN~\cite{rfc4364} and provide the necessary flexibility for adding easily new network changes. The management complexity of MPLS-based VPN solutions hampers the efficiency of VPN provision and maintenance. This is caused by the high number of protocols involved in the control plane such as MP-BGP, LDP, IS-IS and OSPF.

Next generation of DC networks may benefit from the flexibility that Software Defined Network (SDN)~\cite{onf} offers in terms of simplified network management, automation, simplified traffic engineering, etc. In the context of large-scale networks, SDN may enhance the network functionalities in various ways. Firstly, the programmable nature of SDN allows the immediate deployability and adaptability which may alleviate existing problems in DCs such as ARP flooding~\cite{Kim:2008} and long convergence time for learning and updating the network~\cite{zhang2015}. Secondly, SDN offers network abstraction to design network services and the flexibility to deploy an orchestration framework for network provisioning. Thirdly, an SDN-based architecture may benefit from tight integration with public cloud platforms such as OpenStack~\cite{OpenStack} to automatically deploy and flexibly manage various services such as VPNs from a centralized platform. Finally, SDN may collaborate with other frameworks such as the model-driven network to provide a vendor-independent abstraction that translates the set of orders and configurations to a multi-vendor environment. 

In this paper, we propose an SDN-based architecture that flexibly configures and manages EVPN instances (EVIs). This proposed architecture is based on three pillars: 1) EVPN for DC interconnection, 2) model-driven network management, and 3) SDN-based management. The SDN-based architecture employs model-driven network management to automate the deployment of EVIs on DC Provider Edge (PE) routers and bypasses the slow and error-prone tasks of manual EVPN configuration. We extend the MP-BGP module inside the SDN controller to interwork with the MP-BGP control plane (EVIs on the PEs) and the VPNService inside the SDN controller (herein OpenDaylight (ODL)~\cite{Opendaylight}) which automates EVPN configuration using YANG~\cite{rfc6020} and NETCONF. Finally, as a part of our architecture, we also implemented an interface to OpenStack that allows to trigger the orchestrating of the EVPN creation and management through the SDN controller. Moreover, we evaluate the implementation of this architecture in ODL providing an insight into deployment and performance aspects such as scalability and response time.

The remainder of this paper is organized as follows. Section 2 presents the relevant background for our work. Section 3 describes the proposed architecture. In Section 4, we describe and present the results of our experiments and finally in Section 5 we present the conclusions.

\section{Background}
\label{sec:background}
\begin{figure*}[tbh]
\centering{}\includegraphics[width=0.95\textwidth,keepaspectratio]{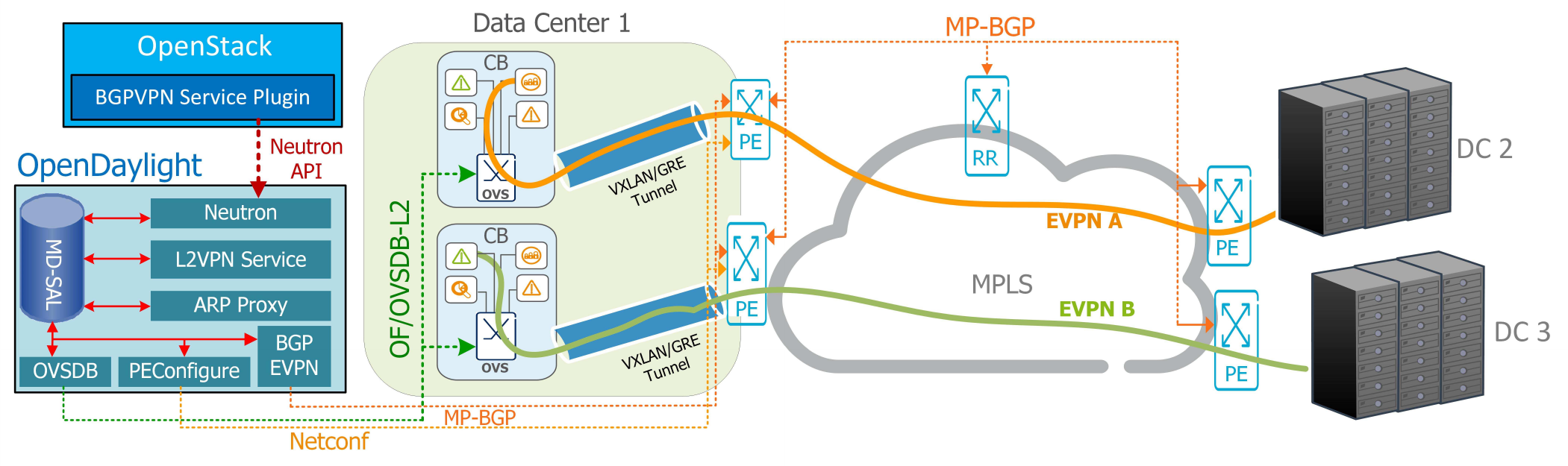}\caption{\label{fig:High-Level-Architecture}High-Level Architecture}
\vspace{-4mm}
\end{figure*}

L2VPN or L3VPN technologies are widely deployed both in DCs and particularly in transport networks to seamlessly interconnect distributed DCs over WAN. In MPLS-based L2VPN solutions, L2  frames are exchanged between different locations over MPLS tunnels. Multiple techniques are used to provide connectivity between remote sites such as Ethernet over MPLS, Point-to-Point (P2P) (e.g., Virtual Private Wire Service) or multipoint-to-multipoint (e.g., VPLS) solutions. Legacy L2VPN solutions do not leverage any signaling mechanism to advertise MAC addresses and flood-and-learn in the data plane is used instead for address learning which imposes extra workload to the network.

EVPN encompasses next generation Ethernet L2VPN solutions and has been designed to handle sophisticated redundancy access scenarios, provide per-flow load balancing, enhance the flexibility and decrease the operational complexity of existing L2VPN solutions. EVPN aligns the well-understood technical and operational principles of IP VPNs to Ethernet services by utilizing MP-BGP in the control plane as a signaling method to advertise addresses which removes the need of traditional flood-and-learn in the data plane. In EVPN, the control and data plane are abstracted and separated. Therefore, several data plane solutions such as MPLS~\cite{rfc7432} and Provider Edge Backbone Bridge~\cite{rfc7623} can be used together with the EVPN control plane. EVPN uses the control plane through extensions of MP-BGP to advertise four type of messages: Ethernet Auto-Discovery, Ethernet Segment, Inclusive Multicast and MAC/IP Advertisement route. For the description and use cases of the aforementioned EVPN messages, we refer the reader to~\cite{rfc7432} for more information.

Model-driven network management automates and accelerates the procedure of creating services through the whole network. In model-driven network management, a data model is used for representing services and configurations together with standard protocols to transmit the modeled data. YANG has clearly positioned itself as the data model language for representing configurations, state data, RPCs and process notifications in a standardized way. Data defined in YANG must be transmitted to a network device using a protocol like NETCONF which allows to install, manipulate, and delete the configuration of network devices based on a server-client connection where the messages are encoded in XML format. Using NETCONF, the administrator pushes the configurations to all devices, validates the configurations and if the validation is successful for all the participants, the administrator commits the changes. Otherwise, the entire configuration can be rolled back.

Not many studies addressed the complexity of VPN management and deployment in the network. Authors in~\cite{Suzuki,vanderPol2016295,7140314} propose different solutions to facilitate the L3VPN deployment and alleviate existing corresponding complexities. However, regarding the L2VPN solution, the number of studies are even less. Authors in \cite{7444836} propose an SDN-based solution to automate VPLS tunnel establishment and reduce the delay of subsequent tunnel establishments between authorized PEs. Authors in \cite{Wood:2011:CDP:1952682.1952699} utilize a central VPN controller to establish the VPLS connection between remote DCs to decrease the VM migration downtime. 

To the best of our knowledge, this is the first work on EVPN deployment automation. Herein, a realistic DC architecture is considered which is equipped with a number of vendor PE routers. Moreover, unlike a number of studies (e.g., \cite{7140314}) we do not propose a new programming language to configure the routers but instead, model the EVPN configuration using YANG as a well-established configuration language. In addition, the controller leverages the standard protocol (NETCONF) to automate the configuration of EVIs on PE routers.



\section{Architecture and Implementation}
\label{sec:architecture}
\subsection{High-level Architecture}
The proposed architecture (see Figure \ref{fig:High-Level-Architecture}) aims to automate and deploy EVPN inside SDN-based DCs is based on:
\begin{itemize}
\item OpenStack: It orchestrates the whole EVPN management process and triggers the association of EVIs to VMs. The OpenStack Neutron API allows OpenStack to interact with ODL for EVPN management.
\item SDN controller (ODL): It creates and manages EVIs on PEs and interacts with remote PEs using MP-BGP protocol.
\item Open VSwtich (OVS): This virtual switch resides inside OpenStack compute nodes, isolates the traffic among different VMs and connects them to the physical network.
\item PE routers:  The PE acts as a gateway for the DCs and supports EVPN and MP-BGP extensions as well as NETCONF and YANG.
\end{itemize}

The routers inside the DC can be OpenFlow-based switches or legacy DC switches. Moreover, we assume that an MPLS-based network is used as data plane to interconnect DCs.

\subsection{Enhanced SDN Functionalities for EVPN\label{subsec:Enhanced-SDN-Functionalities-for-EVPN}}
The SDN controller has been extended to implement the following functionalities in relation to EVPN management:

\textbf{Automate EVPN Deployment}: The administrator sends high-level EVPN deployment commands from OpenStack which uses Neutron extensions to send EVPN configuration to the SDN controller in JSON format. ODL translates the EVPN object into a YANG model and uses NETCONF to send the configuration information to the PEs. 

\textbf{Dynamic Routing Policy (RP) for EVPN}:  \label{subsec:Routing-Policy}
RP defines how the traffic belonging to a specific EVI must be treated. An RP can specify business relationships, traffic characteristics, scalability aspects and security-related policies \cite{1541715}. An RP can be dynamically changed and associated to an EVI. 

\textbf{ARP Suppression}:  An SDN-based architecture may alleviate ARP flooding problem in DC by adding the ARP proxy functionality to the controller. Consequently, when the VM sends an ARP request it is forwarded to the SDN controller. The SDN controller has a table which stores the MAC to IP mappings that it learned locally (from data plane) or from the MAC advertisement messages received through the MP-BGP protocol (see Section \ref{L2VPN}). Consequently, the SDN controller sends the reply to the VM. This process avoids unnecessary flooding operations within the DC.

\textbf{Silent Host}: In a virtualized DC when the VM boots up, it has to announce its existence by generating a Gratuitous ARP (GARP) request which is flooded all over the network and may cause additional traffic load. However, when the VM is not sending GARP request, it leads to the silent host problem where the network entities are not aware of the host which is in operation. SDN controller may learn the creation or migration of new hosts/VMs from the cloud managing platform (e.g., OpenStack) and consequently announce (through EVPN MAC advertisement message) to other network entities to update their tables.

\subsection{SDN Controller Modules\label{subsec:SDN-Controller-Modules}}
The following ODL modules provide the EVPN functionalities:

\subsubsection{Neutron}
The ODL Neutron module has a northbound API for the BGP-VPN service which is in charge of handling the API commands issued by the Neutron client (OpenStack) for the creation, updating and deletion of BGP-based VPN instances. However, this module is extended to handle L2VPN Service and particularly EVPN requests from OpenStack. When the Neutron module receives the requests to create or manage EVIs, it parses the request and prepares an appropriate object for the L2VPN Service.

The DC administrator may send three types of commands: (1) EVPN deployment defining the EVPN parameters such as customer ID\footnote{Customer identical ID that will use the service}, virtual network ID, Service Access Point (SAP) ID\footnote{SAP identifies the customer interface point on a PE router.}, Network IDs and PEs IDs, (2) specify a new RP from OpenStack, and (3) associate an RP to the EVPN. While an EVPN can only be associated to a single RP, an RP can be associated to one or more EVIs.

\subsubsection{L2VPN Service}\label{L2VPN}
This module is an extension of the VPNService module in ODL to support the operation and deployment of L2VPN. The L2VPN Service interacts with other ODL modules such as ODL Neutron, BGP-EVPN, PEConfigure, and OVSDB. The L2VPN Service continuously monitors the RPs, networks, subnets, and ports and immediately reacts upon changes. 
For instance, when a VM is created and associated to one EVI, L2VPN Service advertises the corresponding MAC address to remote PEs if the RP allows it. 
The key responsibilities of this module are the following:

\begin{compactitem}
\item Interoperation with ODL Neutron module to receive the EVPN related commands which are issued from OpenStack.

\item Collect, store and update all parameters related to each EVI and MP-BGP operations e.g., the remote end hosts MAC/IP addresses belonging to such EVPN, MPLS labels, etc.

\item Interact with BGP-EVPN module which receives EVPN control plane messages. When an MP-BGP message concerning EVPN is received, the L2VPN Service stores the received information fetched by the BGP-EVPN module. Additionally, the L2VPN Service determines the execution of EVPN control messages such as MAC advertisement messages and provides the BGP-EVPN module the necessary parameters such as MAC/IP and MPLS label.

\item Provide the EVPN configuration specifications and RP definitions to the PEConfigure module to initialize, update or delete EVPN configuration or RP on each PE belonging to the DC domain.

\item Interchange information with the OVSDB module about the protocols involved in the routing of the traffic within the DC and towards the PEs. It provides information about the VLAN tag, MPLS label or GRE/VXLAN tunnels that must be established from the end hosts (OVS inside the hypervisor) to the end-host or PE. 
\end{compactitem}

\begin{figure}[tbh]
\begin{centering}
\includegraphics[width=0.47\textwidth]{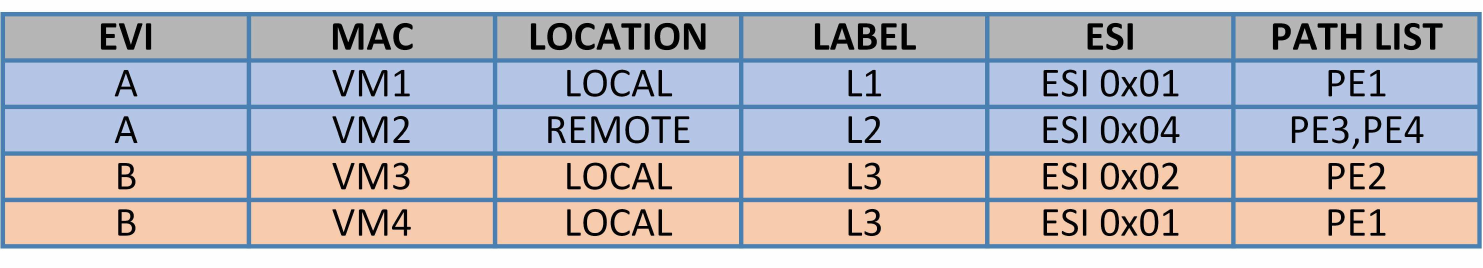}
\par\end{centering}
\caption{\label{fig:MAC-table-data}Overview of the main EVPN table in the L2VPN Service}
\vspace{-5mm}
\end{figure}

\begin{figure}[tbh]
\begin{centering}
\includegraphics[width=0.37\textwidth]{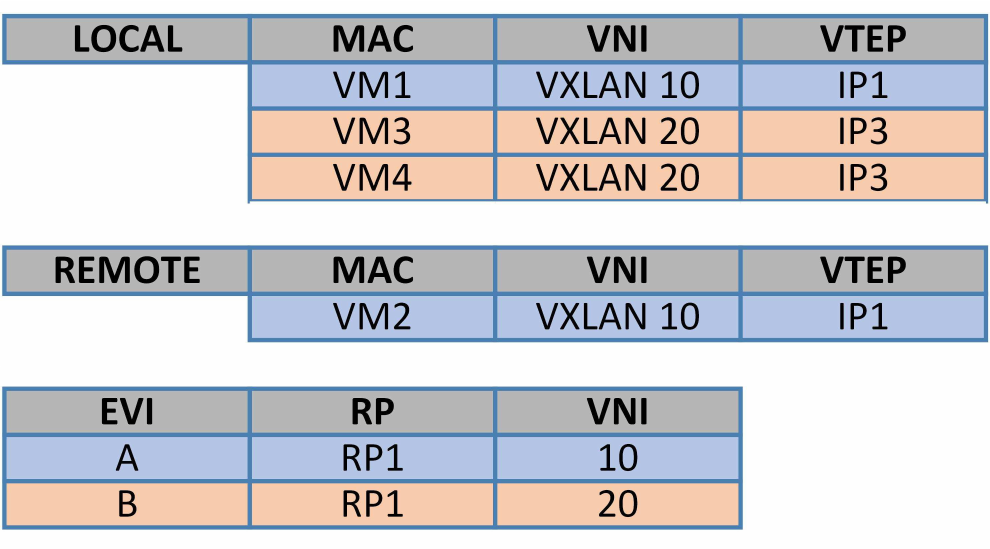}
\par\end{centering}
\caption{\label{fig:Secundary-MAC-table-data}Overview of auxiliary tables in the L2VPN Service}
\end{figure}

In addition to data structures employed by the L2VPN Service to store EVPN and RP parameters, the module has a main local table (Figure \ref{fig:MAC-table-data}) and several auxiliary tables (see Figure \ref{fig:Secundary-MAC-table-data}). The main table contains MAC address information that the controller has locally under its DC domain and learned remotely from remote PEs and their relation with the EVI. Moreover, it stores the MPLS labels associated to each EVI, the Ethernet Segment Identifier (ESI) and the PE(s) of the next hop (path list). The structure of the auxiliary table and the content could differ depending on the protocol used for intra-DC. 

These tables are set up with the information provided by OpenStack, the remote information received via MP-BGP, the ARP Proxy table, and the traffic monitoring of the internal DC. The relations between EVPN and VXLAN segment ID are given by the network administrator which defines the subnets related to a given VXLAN segment ID. The MAC addresses, VMs related to a given EVPN or VXLAN segment ID, and associated RP are also defined by the network administrator via OpenStack. The controller is aware of the PEs participating in a given Virtual Network Identifier (VNI) since this information is distributed using inclusive multicast route over the MPLS network.

\subsubsection{BGP-EVPN}
The existing BGPCEP module in ODL controller was extended to form the BGP-EVPN module in order to parse and serialize MP-BGP messages related to EVPN. It exchanges EVPN related information with the L2VPN Service module and communicates EVPN information to external elements such as PEs and Route Reflectors (RR) using MP-BGP extensions. When the BGP-EVPN module receives a new EVPN MP-BGP control message, it parses the information inside the BGP Network Layer Reachability Information (NLRI) and provides
that information to the L2VPN Service for further actions. Additionally, when the L2VPN Service module needs to advertise or update information belonging to an EVI such as a new MAC address advertisement, it provides address information (e.g., MAC/IP address, MPLS label, and ESI) to the BGP-EVPN module which in turn creates the related MP-BGP message.  

\subsubsection{PEConfigure}
This module prepares the configuration of the PEs for each EVI using the YANG data models and pushes the configuration to PEs using NETCONF protocol. It uses the information provided by the L2VPN Service such as the EVI or its associated RP parameters as well as the ID of the PE. The PEConfigure module parses the given object from the L2VPN Service, extracts the defined parameters, translates them to the PE configuration (according to the YANG specification that the PE uses), and transfers the configuration to the PE. Moreover, when a PE is under the controller domain, this module prepares basic BGP configuration on that PE such as enabling the L2VPN family and defining the neighbors.

\subsubsection{Existing modules}
Besides the deployment or extension of the modules, two main existing ODL modules are used: i) MD-SAL and ii) OVSDB. Modules in ODL leverage the MD-SAL to store objects/configurations and transfer parameters among each other. The data structure and the functionalities of the components are defined using YANG models. OVSDB is a southbound plugin which provides functionalities through the use of OVSDB protocol. This plugin enables the controller to manage the OVS instances running on the hypervisors performing operations such as the creation, manipulation, removing bridges, interfaces, ports and queues in the underlying network. Moreover, this module updates the list of ports and networks in the ODL data store which are used by the L2VPN Service.

\section{Evaluation}
\label{sec:eval}
\subsection{Evaluation Methodology}
We have assessed the controller performance by evaluating the EVPN deployment time and the controller response time to EVPN control plane messages (MAC advertisement). To evaluate the controller response time to EVPN control plane messages, the Bagpipe software router \cite{BaGPipe} is extended to generate the EVPN messages and to stamp all outgoing and incoming packets with the system time. Bagpipe router is configured to operate in three modes:
\begin{enumerate}
\item Burst: Bagpipe is continuously sending a predefined number of EVPN messages within a burst. It waits then for the reply of the sent messages from its peer (ODL).
\item One-by-One: Bagpipe sends one EVPN message and waits for the peer (ODL) reply. As soon as it receives the reply message, Bagpipe sends another message.
\item Single: Bagpipe generates a single EVPN control message.
\end{enumerate}

The experiments run over two 3.2GHz Core i7 processor Intel systems with 8 cores and 16 GB of RAM under Linux 4.4.0 kernel. The first computer hosts the ODL (Beryllium version) and the Alcatel-Lucent virtualized Simulator (vSim). vSim is a virtualization-ready version of Service Router Operating System (SROS) and emulates the control and management plane of an Alcatel-Lucent hardware-based SROS router. vSim version 13.0 R4 is utilized which supports both EVPN and NETCONF protocols. A QEMU instance of the SROS is imported in the GNS3 network emulator.  The second computer hosts the Bagpipe router. The ODL peers with both SROS and Bagpipe routers. The two machines are connected with a 100 Mbps link with 4 ms RTT. All experiments are conducted 5 times to show the average performance of the system in each dataset. To initialize the MD-SAL data store and controller modules to realistic conditions, a number of preliminary messages are sent at the beginning of each experiment. A data logger is added to the controller which stamps the incoming requests that consist of L2VPN, RP, and RP association as well as it stamps the request at the end of their lifecycle.

\subsection{EVPN Deployment Performance}

First, the time required to initialize and deploy an EVI is assessed. The total time is measured as the difference between the initial time that the EVPN creation request is triggered by the administrator via OpenStack and the time instance where the controller receives the EVPN confirmation of its installation in SROS. Recall that the EVPN deployment consists of three steps including 1) EVI creation, 2) RP creation, and 3) RP association. 

We have developed Perl scripts which create L2VPN JSON commands akin to OpenStack outputs, then create the RP and finally associate the RP to the L2VPN instance. These JSON commands are posted to the Neutron interface of the ODL at the appropriate URL. The script waits for 1 second and the same procedure is repeated again. The networks, subnets, and ports are created beforehand and the Perl script randomly assigns network ID(s) to the given L2VPN. We evaluated the controller module performance as we increased the number of EVIs to deploy from 10 to 1000.

\begin{figure}[tbh]
\begin{centering}
\includegraphics[width=0.40\textwidth,keepaspectratio]{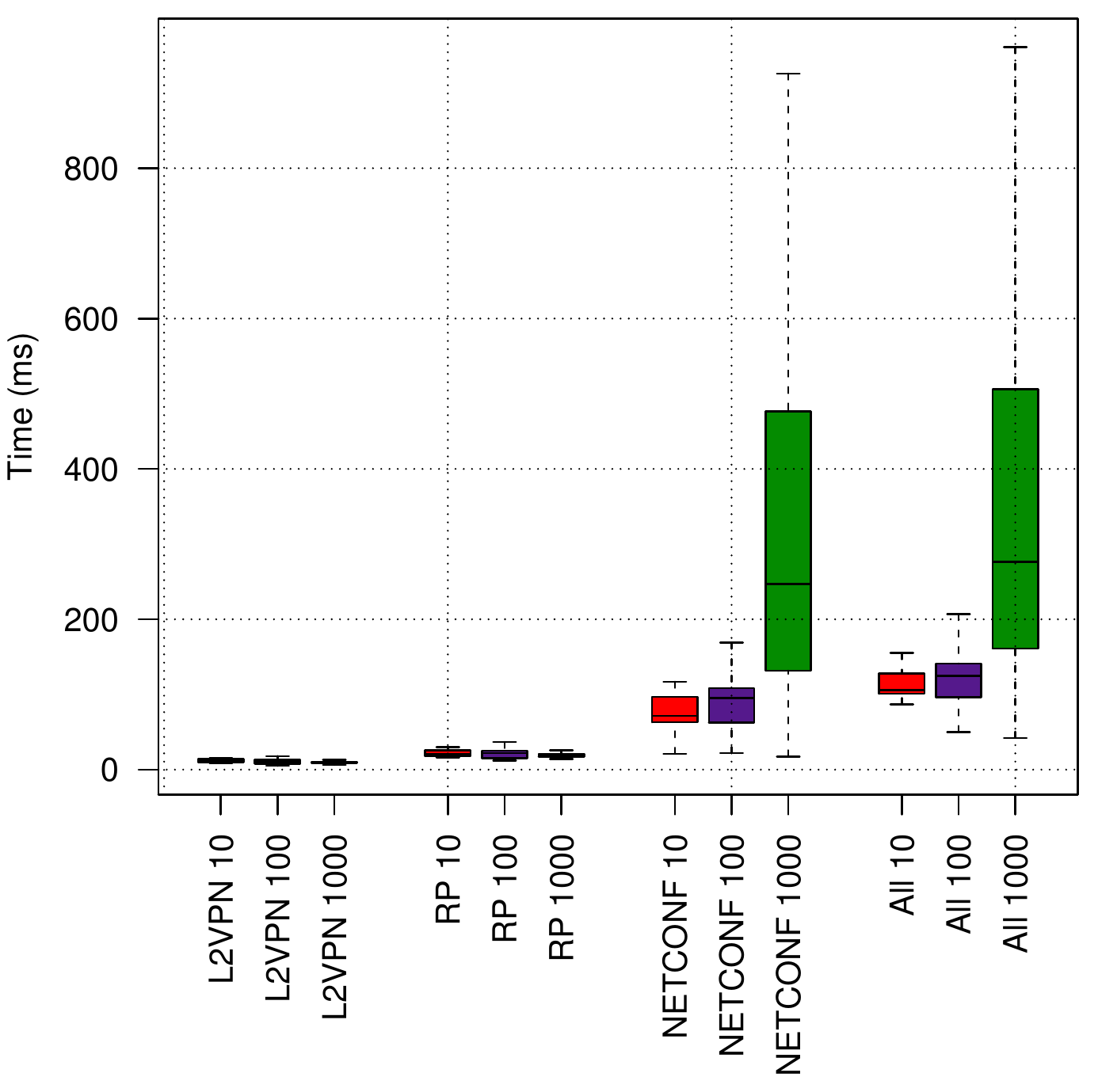}
\par\end{centering}
\caption{\label{fig:EVPN-Deployment-Test}EVPN Deployment Performance Test}
\end{figure}

Figure \ref{fig:EVPN-Deployment-Test} depicts the time consumed by each of the aforementioned steps. The time it takes to create an L2VPN and RP inside the ODL L2VPN Service is relatively small with the average of 9.5 ms and 19.1 ms respectively (for 1000 EVIs). On the other hand, deploying the configuration on the routers is the most time-consuming of the pipeline and the average time is 326.2 ms when there are 1000 EVIs. It is worth to mention that part of our test includes the configuration of EVIs on the virtual router SROS. Consequently, the virtualization may limit the overall performance compared to configuring a real EVPN capable router. Similarly, the control plane CPU allocation in the SROS may also limit the performance of processing the NETCONF messages. The last column in the box plot shows the average total time (sum of L2VPN, RP, and NETCONF) for deploying an EVI. This time is mainly influenced by the PE configuration time and the other operations are almost negligible.

\subsection{Module Performance Test}
In this section, we assess the performance of the SDN controller when the peers (herein Bagpipe) are sending EVPN control plane messages. As we did not have access to a trace that includes real MP-BGP EVPN related messages, the evaluation is performed by instrumenting the BGP-EVPN and the L2VPN Service module using the following scenarios: 

\begin{itemize}
\item The BGP-EVPN module parses the incoming message(s).
\item The BGP-EVPN module passes the parameters to the L2VPN Service.
\item The L2VPN Service updates its local data structure and provides the reply parameter for the new EVPN control plane message.
\item The BGP-EVPN module serializes a new EVPN control plane message with the parameters provided by L2VPN Service and sends it to the peer. 
\end{itemize}

\begin{figure}[tbh]
\begin{centering}
\includegraphics[width=0.45\textwidth,keepaspectratio]{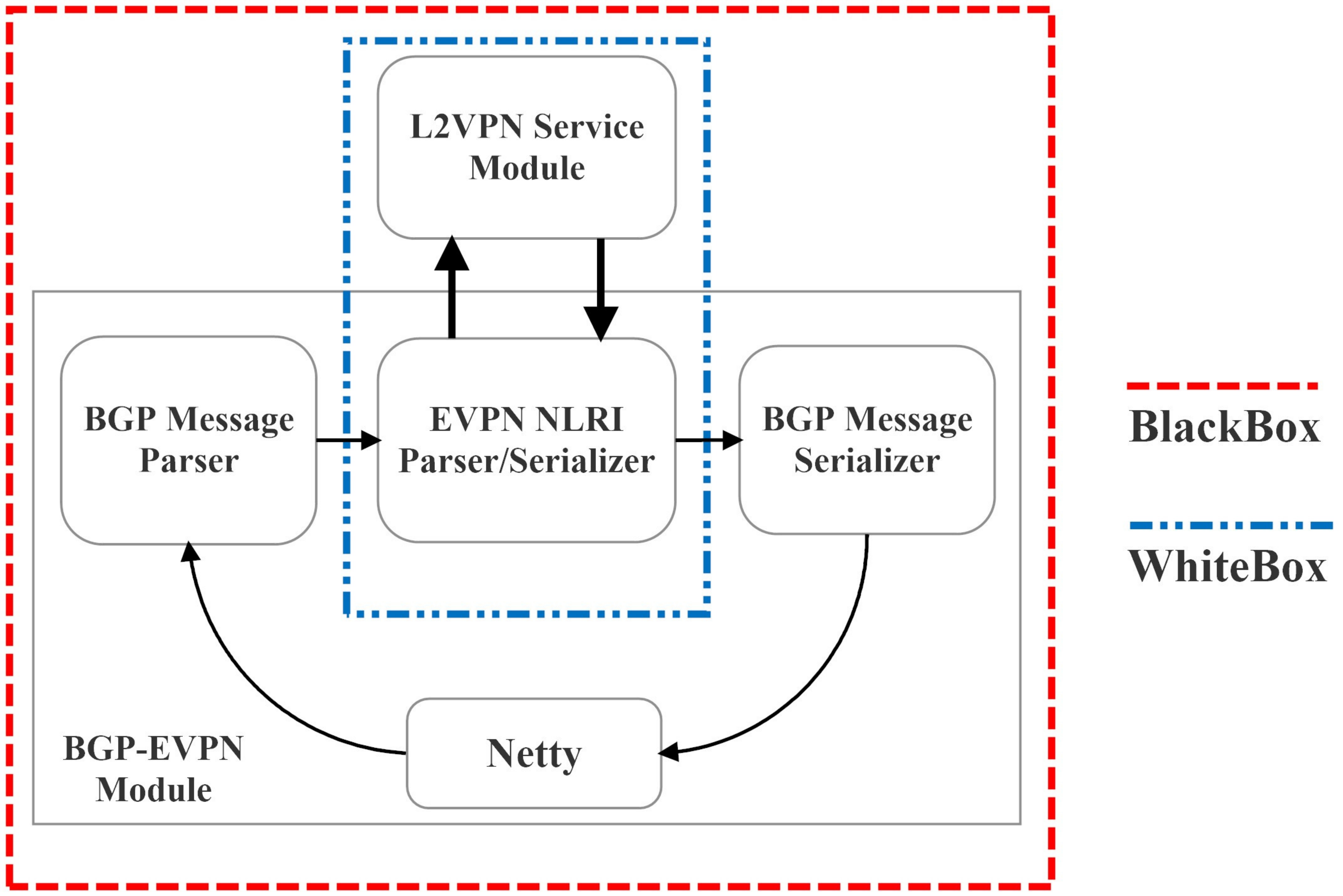}
\par\end{centering}
\caption{\label{fig:Evaluation-Scope}Evaluation Scope}
\end{figure}

The performance of developed modules in the ODL are evaluated in following ways:
\begin{itemize}
\item Whitebox Test (WBT): We measure the time to parse the MP-BGP NLRI segment, till the NLRI segment of the replied message is serialized. In this scenario, the Bagpipe router generates one MAC advertisement message and the ODL stamps the packet at the beginning and at the end of the pipeline.
\item Blackbox Test without Queue (BBT-UQ): Bagpipe operates in One-by-One mode and we measure the additional overhead of the message needed to communicate with ODL. 
\item Blackbox Test with Queue (BBT-Q): Bagpipe operates in burst mode. In this case, ongoing processing in the controller can cause messages to be queued, thus increasing the processing times as measured at the Bagpipe.
\end{itemize}

For the aforementioned BBTs, the controller immediately sends routes back to the Bagpipe to measure the ODL response time. The scope of WBT and BBT are depicted in Figure \ref{fig:Evaluation-Scope}.

\begin{figure}[tbh]
\begin{centering}
\includegraphics[width=0.40\textwidth,keepaspectratio]{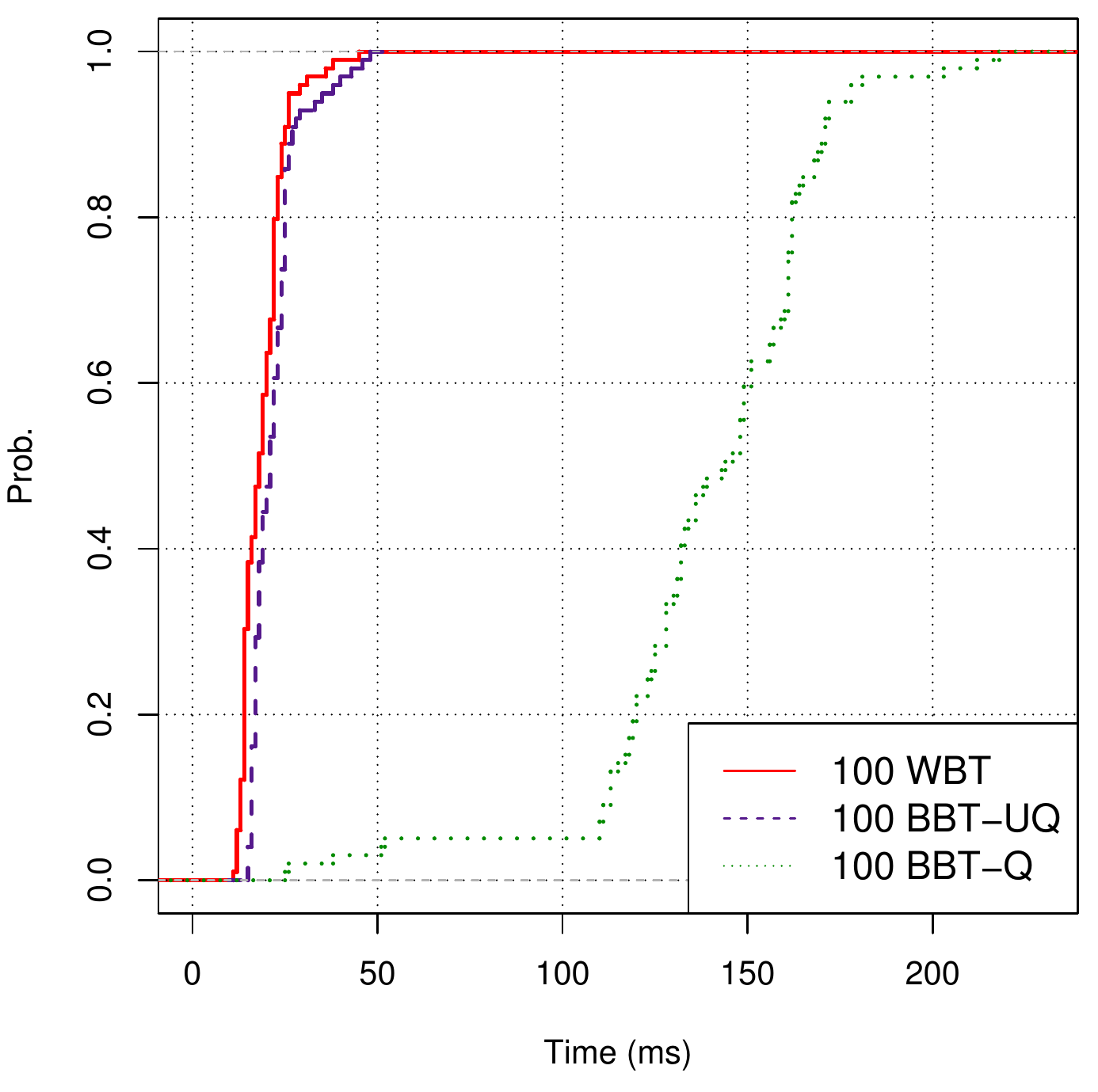}
\par\end{centering}
\caption{\label{fig:Controller-Performance}Controller Performance for Blackbox and Whitebox tests}
\vspace{-3mm}
\end{figure}

Figure \ref{fig:Controller-Performance} depicts the cumulative probability for the WBT and BBT when 100 EVPN messages are exchanged between ODL and Bagpipe. As expected, the message passing adds some overhead to the processing time. However, when there is no queuing effect, the difference between WBT and BBT test is almost negligible. On the other hand, when the Bagpipe sends messages in a burst, the replies reached the Bagpipe with some delays. The reasons for the higher delay are the following: (1) When the ODL observes the session is occupied, it backs off and tries to send messages later which cause additional delay to the ODL responses. Message lifecycle in the ODL begins in the network layer when a message is sent to ODL instance through TCP and ends in the session layer which is handled by Netty (third-party library). (2) Queuing effect starts to be visible inside the ODL to process incoming messages and prepare the reply for each. (3) BGP allows for multiple address prefixes with the same path attributes to be specified in one message. However, this feature causes a delay to send the update messages which are ready since the BGP speaker merges upcoming update messages with the ready ones into one BGP message. 

Moreover, our experiments show that the MD-SAL is the main bottleneck of the pipeline. For instance, in the WBT the EVPN messages are processed and served in 18.86 ms in average, however, almost 50\% (9.39 ms) of this time is consumed in the process of message passing between BGP-EVPN and L2VPN Service module. This bottleneck may be reduced by more tighter integration of data structures inside ODL avoiding the need to pass through the MD-SAL at the expense of less flexibility in reusing those data structures by different modules.

\section{Conclusions and Future Work}
\label{sec:conclusion}
In this paper, we have presented an SDN-based architecture to interconnect various islands of L2 connectivity via a flexible EVPN-based data center interconnection. In our proposed architecture, the SDN controller 1) automates the deployment of EVPN instances using NETCONF and YANG and bypasses the error-prone tasks of EVPN configuration on provider edge routers, 2) manages all EVPN instances and manipulates their configuration according to a given routing policy, and 3) interacts with provider edge routers using EVPN extensions of MP-BGP. Moreover, we have elaborated how this architecture mitigates common problems in a data center such as ARP flooding. We implemented a prototype by extending the common SDN controller OpenDaylight and the open source cloud platform OpenStack. Based on testbed measurements we evaluated the scalability of our solution.

There are numerous next steps that we would like to explore in the future. Regarding routing policies, we intend to evaluate the impact of different load balancing strategies both within and across a data center over the MPLS tunnels. Also, we want to test the scalability of the controller with more realistic VM creation patterns as well as use traces for the scalability tests that contain EVPN messages. 

\section*{Acknowledgment}

Parts of this work has been supported by the Knowledge Foundation Sweden through the profile HITS.

{\small{}\bibliographystyle{./IEEEtran}
\bibliography{refs}
}{\small \par}

\end{document}